\documentclass[12pt]{article}
\usepackage{amssymb,amsmath,latexsym,graphicx, mathrsfs, lscape, amsthm, bm,  hyperref}

\hypersetup{
    colorlinks=true,
    linkcolor=blue,
    citecolor=red,      
    urlcolor=cyan,
}

\numberwithin{equation}{section}

\begin{document}

\begin{titlepage}

    \thispagestyle{empty}

    \vspace{82pt}
    \begin{center}
        { \Huge{\bf Bianchi-IX, Darboux-Halphen and Chazy-Ramanujan}}

        \vspace{18pt}

        {\large{\bf Sumanto Chanda $^\clubsuit$, Partha Guha$^\clubsuit$ and \ Raju Roychowdhury$^{\spadesuit}$}}

        \vspace{15pt}

        {$\clubsuit$ \it S.N. Bose National Centre for Basic Sciences\\
        JD Block, Sector III, Salt Lake, Kolkata 700098, India \\
        \texttt{sumanto12@boson.bose.res.in, partha@boson.bose.res.in}}

        \vspace{15pt}
       
        {$\spadesuit$ \it Instituto de Fisica, Universidade de Sao Paulo,\\
                C. Postal 66318, 05314-970 Sao Paulo, SP, Brazil\\
        \texttt{raju@if.usp.br, raju.roychowdhury@gmail.com}}

       \end{center}

\abstract{Bianchi-IX four metrics are $SU(2)$ invariant solutions of vacuum Einstein equation, for which the connection-wise self-dual case describes the Euler Top, while the curvature-wise self-dual case yields the Ricci flat classical Darboux-Halphen system. It is possible to see such a solution exhibiting Ricci flow. The classical Darboux-Halphen system is a special case of the generalized one that arises from a reduction of the self-dual Yang-Mills equation and the solutions to the related homogeneous quadratic differential equations provide the desired metric. A few integrable and near-integrable dynamical systems related to the Darboux-Halphen system and occurring in the study of Bianchi IX gravitational instanton have been listed as well. We explore in details whether self-duality implies integrability.}

\end{titlepage}

\newpage
\tableofcontents

\newpage

\setcounter{page}{1}
\section{Introduction}

\numberwithin{equation}{section}

Historically, gravitational instantons have been explored with the same motivations similar to gauge instantons: to describe the non-perturbative transitions in quantum gravity, and by analytic continuation producing real-time gravitational backgrounds \cite{GH78, GH79, GH-book}. Dealing with (anti)-self-dual systems makes the task of solving Einsteins equations substantially easier, producing first order equations like the (anti)-self-dual Yang-Mills equations, often related to interesting integrable systems \cite{ACH}. \\

Following applications to homogeneous cosmology, $\mathcal{M}_4$ spaces topologically equivalent to $\mathbb{R} \times \mathcal{M}_3$ of Bianchi type have been extensively explored. For a pedagogical introduction to various cosmological models look at the Lecture Notes by Ellis et.al.\cite{lecture}. Out of all Bianchi type models of classes I - IX with vanishing cosmological constant, only Bianchi-IX has been found to exhibit a relationship with quasimodular forms \cite{MM1, MM2}. Modular forms in physics are a consequence of duality properties, resulting either from an invariance or a relationship between two distinct theories. In the past 30 years, modular and quasimodular forms have emerged mostly in the study of gravity and string theory \cite{MPPV}. Furthermore, we must note that the Bianchi-IX model is a controversial system (possessing both integrable and non-integrable aspects).

A debate followed between various authors involving doubts over statements regarding the chaotic nature of Bianchi-IX dynamics. They simultaneously expressed their opinion that the model might as well be a classical integrable system (In Liovilles sense) \cite{HBC}. Thus Bianchi-IX cosmological model is a good ``thought laboratory'' for testing theories in order to understand various concepts of integrability. \\

Since the late 70s, many systems were developed starting with self-dual Yang-Mills system after reductions, leading to the belief that all integrable systems were obtained from similar such reductions \cite{ACH, PMPetro1, PMPetro2}. The generalized Darboux-Halphen system is one such system, heavily studied over the recent years, although in its classical form the relationship with Bianchi-IX metric was established rather recently in the 90s. Such a connection proves to be of use in constructing various interesting Bianchi IX solutions \cite{GibbonsCQG} or their applications e.g. in the study of scattering of $SU(2)$ BPS monopoles \cite{PMPetro1, AC}. The Darboux-Halphen system also exhibits Ricci flow that describes the evolution of $SU(2)$-homogeneous 3D geometries and can be seen as reflection of hidden symmetrry of hyperbolic monopole motion \cite{GibWar}. \\

The plan of this article is as follows: In Sect.2 we will first perform a geometric analysis of the Bianchi-IX metric, directly exploring both connection-wise and curvature-wise self-dual cases, bringing us to the Eule-Top and classical Darboux-Halphen cases respectively, followed by computation of the general form of the curvature components. Then in the next Sect.3 we will start with the self-dual Yang-Mills equation  and by reduction see how it gives rise to the generalized Darboux-Halphen system, which has a close relation to the Euler-Top system. This will be followed by an elaboration on solutions that can be obtained for such a system. A brief note will be made as to how the classical case arises from the generalized one and also explore why we cannot always find a metric that gives rise to the generalized system. We will  proceed to list solutions to the various first-order differential equations involved with the system, that effectively yield the metric corresponding to the classical system and possibly the generalized one as well. This will be followed by a derivation and examination of the Ricci flow equations in Sect.4. We will see in Sect.5 how the Chazy equation emerges from the classical Darboux-Halphen system, a result of curvature-wise self-duality, as well how othes like the Ramanujan and Ramamani systems are related to it. This will be followed up by a detailed analysis of integrability of the Bianchi-IX in Sect.6 to see if self-duality implies integrability. Fianlly we conclude in Sect.7 and try to look for future directions. \\

\textbf{Note added}: Near the completion of this work, the paper \cite{FFM} appeared on the arXiv. The authors, among other things, also address the question of arithmetics and integrability of Bianchi IX gravitational instantons, which have been explored by imposing the self-duality condition on triaxial Bianchi IX metrics and by employing a time-dependent conformal factor. We comment more about this in the final section at the end of the paper.

\section{Geometric analysis}

The Bianchi-IX metric is a general setup for 4D Euclidean spherically symmetric metrics. Under certain settings of its curvature-wise anti-self-dual case, becomes the Taub-NUT. Naturally, the analysis of its connection and curvature follows the same way as in \cite{CGR1, CGR2}. \\

The metric is written as:

\begin{equation}
\label{bianchiix} ds^2 = \big[ c_1 (r) c_2 (r) c_3 (r) \big]^2 dr^2 + c_1^2 (r) \sigma_1^2 + c_2^2 (r) \sigma_2^2 + c_3^2 (r) \sigma_3^2
\end{equation}

where the variables $\sigma_i$ obey the structure equation:

\begin{equation}
\label{struc} d \sigma^i = - {\varepsilon^i}_{jk} \ \sigma^j \wedge \sigma^k \hspace{1cm} \text{where} \hspace{0.75cm} \sigma^i = - \frac1{r^2} \eta^i_{\mu \nu} x^\mu dx^\nu
\end{equation}

and $i, j, k$ are permutations of the indices $1, 2, 3$. \\

Those solutions that are (anti-) self-dual fall into 2 categories:

\begin{enumerate}

\item connection wise self-dual

\item curvature wise self-dual

\end{enumerate}

we shall now uncover the systems characterising each category respectively.

\numberwithin{equation}{subsection}

\subsection{Connection wise self duality - the Lagrange system}

First we shall compute the spin connections in the same manner as for the Taub-NUT \cite{CGR2}. We can describe the tetrads as:

\vspace{-0.5cm}
\begin{equation}
\begin{split}
e^0 = c_0 (r) dr \hspace{1cm} e^i &= c_i (r) \sigma^i, \hspace{0.5cm} i = 1, 2, 3 \\
\text{where} \hspace{1cm} c_0 (r) &= c_1 (r) c_2 (r) c_3 (r)
\end{split}
\end{equation}

Obviously, $e^0$ produces no connections $( d e^0 = 0 )$. However, for the remaining three:

\begin{equation}
\label{deriv} d e^i = - \frac{\partial_r c_i}{c_0} \ \sigma^i \wedge e^0 - \bigg\{ - {\varepsilon^i}_{jk} \frac{ c_i^2 + c_j^2 - c_k^2 }{2 c_i c_j} \ \sigma^k \wedge e^j  - {\varepsilon^i}_{kj} \frac{ c_i^2 + c_k^2 - c_j^2 }{2 c_i c_k} \ \sigma^j \wedge e^k \bigg\}
\end{equation}

Under torsion-free condition the 1st Cartan structure equation $( d e^i = - {\omega^i}_j \wedge e^j )$ gives us the following spin connections:

\vspace{-0.25cm}
\begin{equation}
{\omega^i}_0 = \frac{\partial_r c_i}{c_0} \ \sigma^i \hspace{2cm} {\omega^i}_j = - {\varepsilon^i}_{jk} \frac{ c_i^2 + c_j^2 - c_k^2 }{2 c_i c_j} \ \sigma^k
\end{equation}

This elaborate form for the components of the spin connections make its anti-symmetric nature evident. If we only consider those cases of (anti-) self-duality, we will have

\[ \omega_{0i} = \pm \frac12 {\varepsilon_{0i}}^{jk} \omega_{jk} = \pm \omega_{jk} \hspace{1cm} \Rightarrow \hspace{1cm} 2 \frac{\partial_r c_i}{c_i} = \mp \varepsilon_{jki} \big( c_j^2 + c_k^2 - c_i^2 \big) \]

\begin{equation}
\label{consd} \therefore \hspace{1cm} \partial_r \big( \ln c_i^2 \big) = \mp \varepsilon_{jki} \big( c_j^2 + c_k^2 - c_i^2 \big)
\end{equation}

One may suppose that we must parametrize the LHS to match the linear form of the RHS in the equation above. Essentially, the derivative operator aside, $c_i^2$ must be parametrized such that

\[ \ln c_i^2 = \ln \Omega_j + \ln \Omega_k - \ln \Omega_i = \ln \bigg( \frac{\Omega_j \Omega_k}{\Omega_i} \bigg) \]

\vspace{-0.25cm}
\begin{equation}
\label{pmtr} \therefore \hspace{1cm} ( c_i )^2 = \frac{\Omega_j \Omega_k}{\Omega_i} \hspace{1cm} \Rightarrow \hspace{1cm} \Omega_i = c_j c_k
\end{equation}

which enable us to decouple the individual parameters into their own equations turning into simpler expressions. This allows us to continue our analysis from (\ref{consd}) as

\[ \partial_r \bigg[ \ln \bigg( \frac{\Omega_j \Omega_k}{\Omega_i} \bigg) \bigg] = \frac{\dot{\Omega}_j}{\Omega_j} + \frac{\dot{\Omega}_k}{\Omega_k} - \frac{\dot{\Omega}_i}{\Omega_i} = \mp \varepsilon_{jki} \bigg( \frac{\Omega_k \Omega_i}{\Omega_j} + \frac{\Omega_i \Omega_j}{\Omega_k} - \frac{\Omega_j \Omega_k}{\Omega_i} \bigg) \]

Adding up with a similar expression for $2 \ \partial_r \big( \ln c_j \big) = \mp \varepsilon_{kij} \big( c_i^2 + c_k^2 - c_j^2 \big)$, we get

\[ \frac{\dot{\Omega}_i}{\Omega_i} + \frac{\dot{\Omega}_k}{\Omega_k} - \frac{\dot{\Omega}_j}{\Omega_j} = \mp \varepsilon_{kij} \bigg( \frac{\Omega_j \Omega_k}{\Omega_i} + \frac{\Omega_i \Omega_j}{\Omega_k} - \frac{\Omega_k \Omega_i}{\Omega_j} \bigg) \]

we will find that self-dual cases of the Bianchi-IX (keeping in mind that $\varepsilon_{jki} = \varepsilon_{ijk} = \varepsilon_{kij} = 1$) metric gives us the Lagrange (Euler-top) system

\vspace{-0.5cm}
\[ \begin{split}
\frac{\dot{\Omega}_j}{\Omega_j} + \frac{\dot{\Omega}_k}{\Omega_k} - \frac{\dot{\Omega}_i}{\Omega_i} &= \mp \varepsilon_{jki} \bigg( \frac{\Omega_k \Omega_i}{\Omega_j} + \frac{\Omega_i \Omega_j}{\Omega_k} - \frac{\Omega_j \Omega_k}{\Omega_i} \bigg) \\
& \hspace{0.5cm} + \\
\frac{\dot{\Omega}_i}{\Omega_i} + \frac{\dot{\Omega}_k}{\Omega_k} - \frac{\dot{\Omega}_j}{\Omega_j} &= \mp \varepsilon_{kij} \bigg( \frac{\Omega_j \Omega_k}{\Omega_i} + \frac{\Omega_i \Omega_j}{\Omega_k} - \frac{\Omega_k \Omega_i}{\Omega_j} \bigg) \\
& \hspace{0.5cm} \big{\downarrow}
\end{split} \]

\begin{equation}
\hspace{1cm} \therefore \hspace{1cm} 2 \frac{\dot{\Omega}_k}{\Omega_k} = \mp 2 \frac{\Omega_i \Omega_j}{\Omega_k} \hspace{1cm} \Rightarrow \hspace{1cm} \boxed{\dot{\Omega}_k = \mp \Omega_i \Omega_j}
\end{equation}

where throughout derivative (denoted by dot) is taken with respect to $r$.

\subsection{Curvature wise self-duality - Classical Darboux-Halphen system}

Since we have already covered connection-wise self duality, let us explore a stronger version known as curvature-wise self-duality. This emphasizes and expands upon the property of self-duality, generalizing it beyond connection 1-forms. This means that curvature-wise self-duality does not invalidate, rather implies connection-wise self-duality \cite{ETH}, and hence part of the dynamical system derived from this should have the same form as the Lagrange system. \\

 The Cartan-structure equation for Riemann curvature is

\begin{equation}
\label{cartan2} R_{ij} = d \omega_{ij} + \omega_{im} \wedge {\omega^m}_j
\end{equation}

The self-duality of curvature demands that

\begin{equation}
\label{curvsd} R_{0i} = \frac12 { \varepsilon_{0i} }^{jk} R_{jk} = R_{jk}
\end{equation}

Now, for the LHS and RHS of (\ref{curvsd}), we have

\vspace{-0.5cm}
\begin{align}
\label{lhs} \text{LHS} \hspace{1cm} R_{0i} &= d \omega_{0i} + \omega_{0j} \wedge {\omega^j}_i + \omega_{0k} \wedge {\omega^k}_i \\ \nonumber \\
\label{rhs} \text{RHS} \hspace{1cm} R_{jk} &= d \omega_{jk} + \omega_{j0} \wedge {\omega^0}_k + \omega_{ji} \wedge {\omega^i}_k \nonumber \\
&= d \omega_{jk} - \omega_{0j} \wedge {\omega^0}_k - \omega_{ji} \wedge \omega_{ki}
\end{align}

Now, for (anti-) self-duality of the connection forms, as employed in the previous section, we shall be able to eliminate some of the later terms of (\ref{lhs}) and (\ref{rhs}), since

\begin{equation}
\label{condual} \omega_{ij} = \pm \frac12 { \varepsilon_{ij} }^{k0} \omega_{k0} = \mp \omega_{k0} \hspace{1cm} \Rightarrow \hspace{1cm} \omega_{ji} \pm \omega_{0k} = 0
\end{equation}

This leaves us with the equation shown below and its solution that follows, adapted from the previous subsection.

\vspace{-0.35cm}
\begin{equation}
d \omega_{0i} = \pm d \omega_{jk} 
\end{equation}


\vspace{-0.5cm}
\[ \begin{split}
\Rightarrow \hspace{1cm} &\partial_r \bigg( \frac{\partial_r c_i}{c_0} \bigg) dr \wedge \sigma^i + \frac{\partial_r c_i}{c_0} d \sigma^i = \mp \partial_r \bigg( \frac{ c_j^2 + c_k^2 - c_i^2 }{2 c_j c_k}  \bigg) dr \wedge \sigma^i \mp \frac{ c_j^2 + c_k^2 - c_i^2 }{2 c_j c_k} \ d \sigma^i \\
\Rightarrow \hspace{1cm} &\partial_r \bigg( \frac{\partial_r c_i}{c_0} \bigg) = \mp \partial_r \bigg( \frac{ c_j^2 + c_k^2 - c_i^2 }{2 c_j c_k}  \bigg) \hspace{1cm} \Rightarrow \hspace{1cm} \frac{\partial_r c_i}{c_0} = \mp \varepsilon_{jki} \frac{ c_j^2 + c_k^2 - c_i^2 }{2 c_j c_k} + \lambda_{jk} \\ \\
& \hspace{1cm} \Rightarrow \hspace{1cm} 2 \partial_r  \big( \ln c_i \big) = \mp \big( c_j^2 + c_k^2 - c_i^2 \big) + 2 \lambda_{jk} c_j c_k
\end{split} \]

Thus, as before, upon parametrization we shall have

\vspace{-0.5cm}
\begin{align}
\frac{\dot{\Omega}_j}{\Omega_j} + \frac{\dot{\Omega}_k}{\Omega_k} - \frac{\dot{\Omega}_i}{\Omega_i} &= \mp \bigg( \frac{\Omega_k \Omega_i}{\Omega_j} + \frac{\Omega_i \Omega_j}{\Omega_k} - \frac{\Omega_j \Omega_k}{\Omega_i} \bigg) + 2 \lambda_{jk} \Omega_i \\ \nonumber \\
\frac{\dot{\Omega}_i}{\Omega_i} + \frac{\dot{\Omega}_k}{\Omega_k} - \frac{\dot{\Omega}_j}{\Omega_j} &= \mp \bigg( \frac{\Omega_j \Omega_k}{\Omega_i} + \frac{\Omega_i \Omega_j}{\Omega_k} - \frac{\Omega_k \Omega_i}{\Omega_j} \bigg) + 2 \lambda_{ik} \Omega_j 
\end{align}

Adding up these two results, just like before, will now give us

\vspace{-0.25cm}
\begin{equation}
\label{dhalpsys} \dot{\Omega}_k = \mp \Omega_i \Omega_j + \lambda_{jk} \Omega_k \Omega_i + \lambda_{ik} \Omega_k \Omega_j
\end{equation}

where setting $\lambda_{ij} = - 1 \hspace{0.25cm} \forall \hspace{0.25cm} i, j$ in (\ref{dhalpsys}) for anti-self-duality proceeds to give us the classical Darboux-Halphen system

\begin{equation}
\therefore \hspace{1cm} \boxed{\dot{\Omega}_k = \Omega_i \Omega_j - \Omega_k \big( \Omega_i + \Omega_j \big)}
\end{equation}

Thus, we can see that the curvature-wise self-duality extends upon the characteristic system of the connection-wise self-duality, making the Darboux-Halphen system a suitable candidate for further development beyond the Lagrange system. Clearly, the first term has included the dynamical aspect of the Lagrange system, as the property of self-duality of the connection 1-forms being preserved, aside from an additive constant involved and was extended to their exterior derivatives. Needless to say, connection-wise self-duality must precede curvature-wise self-duality, and the latter is not possible without ensuring the former.

\subsection{Self-dual curvature components}

So far, we have managed to study a great deal about the Bianchi-IX geometry, without confronting the work of extracting the curvature components. Now, in this subsection, we will proceed to do exactly that, using the imposed (anti-) self duality properties at our disposal to make our job easier. But first, we shall prove and later in this case confirm that all curvature-wise self-dual manifolds are Ricci-flat. \\

We recall from \cite{CGR1} that the Riemann curvature tensor for (anti-)self-dual metrics on the vierbein space can be written as:

\begin{equation}
\label{sdcur} R_{abcd} = \mathcal{G}_{ij} (\vec{x}) \eta^{(\pm)i}_{ab} \eta^{(\pm)j}_{cd} \hspace{1cm} i, j = 1, 2, 3; \hspace{0.5cm} a, b, c, d = 0, 1, 2, 3
\end{equation}

This means the Ricci tensor for Euclidean vierbein space is given by

\begin{equation}
\mathbb{R}_{ac} = \delta^{bd} R_{abcd} = \delta^{bd} \mathcal{G}_{ij} (\vec{x}) \eta^i_{ab} \eta^j_{cd} = \mathcal{G}_{ij} (\vec{x}) \delta^{ij} \delta_{ac} = \Big( \text{Tr} \big[ \mathcal{G} (\vec{x}) \big] \Big) \delta_{ac}
\end{equation}

Clearly, the above result implies that the Ricci tensor has only diagonal elements, which allows us to demonstrate that

\[ \mathbb{R}_{aa} = R_{abab} +  R_{acac} +  R_{adad} \xrightarrow{\text{self-duality}} \pm \big( R_{abcd} +  R_{acdb} +  R_{adbc} \big)  \xrightarrow{\text{Bianchi identity}} 0 \]

\vspace{-0.25cm}
\begin{equation}
\therefore \hspace{1cm} \boxed{ \mathbb{R}_{aa} = 0 } \hspace{1cm} \Rightarrow \hspace{1cm} \boxed{ \text{Tr} \big[ \mathcal{G} (\vec{x}) \big] = 0}
\end{equation}

Showing that curvature-wise self-dual manifolds are undoubtedly Ricci-flat.

\[ \boxed{\boxed{\text{Self-duality} \hspace{0.5cm} \Longrightarrow \hspace{0.5cm} \text{Ricci-flatness}}} \]

Returning to the original co-ordinates, we have:

\begin{equation}
\label{ric} \mathbb{R}_{ac} = \big( \mathcal{G}_{ij} (\vec{x}) \delta^{ij} \big) \big( \delta_{ac} {e^a_\mu} {e^c_\nu} \big) = \text{Tr} \big[ \mathcal{G} (\vec{x}) \big] g_{\mu \nu} ( \vec{x} )
\end{equation}

But, for a more thorough analysis, it would be better to directly obtain all the curvature components for detailed examination. This can be easily done as we already have the general formula for all the connection components. The results are made easier by keeping the self-duality of the connection forms (\ref{condual}) in mind.

\vspace{-0.25cm}
\[ \begin{split}
R_{0i} &= d \omega_{0i} + \omega_{0j} \wedge \omega_{ji} + \omega_{0k} \wedge \omega_{ki} \nonumber \\
&= d \omega_{0i} + \omega_{0j} \wedge \omega_{0k} - \omega_{0k} \wedge \omega_{0j} \nonumber = d \omega_{0i} + 2 \omega_{0j} \wedge \omega_{0k}
\end{split} \]

\begin{equation}
\therefore \hspace{1cm} R_{0i} = \underbrace{\frac1{c_0 c_i} \bigg( \frac{c_i '}{c_0} \bigg)'}_{R_{0i0i}} e^0 \wedge e^i - \underbrace{\frac1{c_j c_k} \bigg[ \frac{c_i '}{c_0} - 2 \frac{\big( c_j ' \big) \big( c_k ' \big)}{c_0^2} \bigg]}_{- R_{0ijk}} e^j \wedge e^k
\end{equation}

Now curvature wise anti-self-duality means

\begin{equation}
R_{0i0i} = - R_{0ijk} = - R_{jk0i} = R_{jkjk}
\end{equation}

Demanding curvature wise anti-self-duality gives us the differential equation

\begin{equation}
\frac1{c_0 c_i} \bigg( \frac{c_i '}{c_0} \bigg)' = \frac1{c_j c_k} \bigg[ \frac{c_i '}{c_0} - 2 \frac{\big( c_j ' \big) \big( c_k ' \big)}{c_0^2} \bigg]
\end{equation}

Since we have connection wise anti-self-duality rule (\ref{consd}), we can say

\vspace{-0.5cm}
\begin{align}
R_{0i} = \frac{\varepsilon_{ijk}}{c_0 c_i} &\bigg( \frac{c_j^2 + c_k^2 - c_i^2}{2 c_j c_k} \bigg)' e^0 \wedge e^i - \frac{\varepsilon_{ijk}}{c_j c_k} \bigg[ \frac{c_j^2 + c_k^2 - c_i^2}{2 c_j c_k} - 2 \bigg( \frac{c_k^2 + c_i^2 - c_j^2}{2 c_k c_i} \bigg) \bigg( \frac{c_i^2 + c_j^2 - c_k^2}{2 c_i c_j} \bigg) \bigg] e^j \wedge e^k \nonumber \\ \nonumber \\
\therefore \hspace{0.5cm} R_{0i} &= \underbrace{\frac{\varepsilon_{ijk}}{c_0 c_i} \bigg( \frac{c_j^2 + c_k^2 - c_i^2}{2 c_j c_k} \bigg)'}_{R_{0i0i}} e^0 \wedge e^i - \underbrace{\varepsilon_{ijk} \frac{c_i^2 \big( c_j^2 + c_k^2 - c_i^2 \big) - c_i^4 + \big( c_j^2 - c_k^2 \big)^2}{2 c_0^2}}_{- R_{0ijk}} e^j \wedge e^k
\end{align}

Due to curvature wise anti-self-duality being considered, we should have:

\begin{equation}
R_{0i0i} = - R_{0ijk} = \varepsilon_{ijk} \frac{\big( c_i^2 + c_j^2 + c_k^2 \big) \big( c_j^2 + c_k^2 \big) - 2 c_i^4 - 4 c_j^2 c_k^2}{2 c_0^2}
\end{equation}

Thus, we can say that the curvature 2-form is given by

\begin{equation}
\boxed{R_{ab} = \sum_{i = 1}^3 \varepsilon_{ijk} \frac{\big( c_i^2 + c_j^2 + c_k^2 \big) \big( c_j^2 + c_k^2 \big) - 2 c_i^4 - 4 c_j^2 c_k^2}{2 c_0^2} \ \bar{\eta}^i_{ab} \bar{\eta}^i_{cd} e^c \wedge e^d}
\end{equation}

which on comparison with (\ref{sdcur}) tells us that

\vspace{-0.5cm}
\begin{align}
\mathcal{G}_{il} (\vec{x}) &= \varepsilon_{ijk} \frac{\big( c_i^2 + c_j^2 + c_k^2 \big) \big( c_j^2 + c_k^2 \big) - 2 c_i^4 - 4 c_j^2 c_k^2}{2 c_0^2} \delta_{il} \\
\text{Tr} \big[ \mathcal{G} \big] &= \frac{2 \big( c_i^2 + c_j^2 + c_k^2 \big)^2 - 2 \big( c_i^2 + c_j^2 + c_k^2 \big)^2}{2 c_0^2} = 0
\end{align}

Thus, the Ricci tensor, and consequently scalar on vierbein space is given by

\vspace{-0.25cm}
\begin{equation}
\boxed{\mathbb{R}_{ab} = 0, \hspace{1cm} \mathbb{R} = 0}
\end{equation}

confirming what was already proven previously.

\subsection{Special case: The Taub-NUT}

The Taub-NUT is a special case of the Bianchi-IX metric connected to the classical Darboux-Halphen system. It is an exact solution of Einstein's equations, found by Abraham Huskel Taub (1951) and extended to a larger manifold by E. Newman, T. Unti and L. Tamburino (1963), whose names are included in the name of the metric Taub-NUT. It exhibits anti-self-duality as can be seen from its curvature and $SU(2)$ gauge fields. Recently, we have conducted a detailed study of the Taub-NUT \cite{CGR2} where we have explored its self-duality, geometric properties, conserved quantities and Killing tensors. \\

The condition to set to obtain this metric is $\Omega_1 = \Omega_2 = \Omega \neq \Omega_3$. This consequently makes $c^2_1 = c^2_2 = c^2 = \Omega_3$ and converts the Bianchi-IX metric into the following form:

\begin{equation}
ds^2 = \Omega^2 \Omega_3 \ dr^2 + \Omega_3 \big( \sigma_1^2 + \sigma_2^2 \big) + \frac{\Omega^2}{\Omega_3} \sigma_3^2
\end{equation}

Naturally, we would have to define the parameters and rescale the radial co-ordinate:

\begin{equation}
dr = \frac1{2m} \frac{d \tilde{r}}{\Omega^2} \hspace{0.5cm} \Rightarrow \hspace{0.5cm} ds^2 = \frac{\Omega_3}{\Omega^2} \ \frac{d \tilde{r}^2}{4 m^2} + \Omega_3 \big( \sigma_1^2 + \sigma_2^2 \big) + \frac{\Omega^2}{\Omega_3} \sigma_3^2
\end{equation}

Finally, we can define the parameters as

\begin{equation}
\Omega = \frac{\tilde{r} - m}{2 m} \hspace{1cm} \Omega_3 = \frac{\tilde{r}^2 - m^2}{4 m^2}
\end{equation}

which means that the rescaling equation is

\begin{equation}
dr = \frac{2m}{ ( \tilde{r} - m )^2 } \ d \tilde{r} \hspace{1cm} \Rightarrow \hspace{1cm} r = k - \frac{2m}{ \tilde{r} - m }
\end{equation}


This finally leads to Taub-NUT metric up to a rescaling conformal factor, 

\vspace{-0.25cm}
\begin{equation}
ds^2 = \frac{\tilde{r} + m}{\tilde{r} - m} d \tilde{r}^2 + 4m^2 \frac{\tilde{r} - m}{\tilde{r} + m} \big( d \psi + \cos \theta \ d \phi \big)^2 + \big( \tilde{r}^2 - m^2 \big) \big( d \theta^2 + \sin^2 \theta \ d\phi^2 \big)
\end{equation}

\section{The generalized Darboux-Halphen system}

So far, we have dealt with the classical Darboux-Halphen system that arises as the connection-wise self-dual case of the Bianchi-IX metric. Now we shall proceed to derive a generalized version of this system starting from a different configuration as described in \cite{ACH, dhsstruc}.

\subsection{Reduction of the SDYM equation}

While the classical Darboux-Halphen system is an artifact of connection-wise self-duality of Bianchi-IX, this time we shall start by considering self-dual Yang-Mills equation and execute a reduction process on it.

\begin{equation}
\label{sdym} F_{ab} = - \frac12 {\varepsilon_{ab}}^{cd} F_{cd} \hspace{2cm} F_{ab} = \partial_a A_b - \partial_b A_a - \big[ A_a, A_b \big]
\end{equation}

If we set $A_0 = 0$ and all $A_i = A_i (t)$ only, then (\ref{sdym}) becomes the Nahm equation \cite{Nahm, Hitch, Donald}

\begin{equation}
\label{nahm} F_{0i} = \dot{A}_i \hspace{2cm} F_{ij} = - \big[ A_i, A_j \big] \hspace{1cm} \Rightarrow \hspace{1cm} \dot{A}_i = \frac12 {\varepsilon_i}^{jk} \big[ A_j, A_k \big]
\end{equation}

Now, the $A_i$s are functions from $\mathbb{R}^4$ to a Lie algebra $\mathfrak{g}$ given by

\begin{equation}
\label{fld} A_i = - M_{ij} (t) O_{jk} X_k
\end{equation}

where $O_{ij}$ is an SO(3) matrix, and $X_i$ are the generators of $\mathfrak{sdiff}(S^3)$ satisfying the relation $\big[ X_i, X_j \big] = \varepsilon_{ijk} X_k$. The matrix $M_{ij}$ is given as a sum of  symmetric components $M_s$ and anti-symmetric components $M_a$

\vspace{-0.5cm}
\begin{align} \displaybreak[0]
\label{mat} M &= M_s + M_a = P (d + a) P^{-1} \hspace{2cm} M_a = {\varepsilon_{ij}}^k \tau_k \\ \nonumber \\
\text{where } \hspace{1cm} &M_s = \left({\begin{array}{ccc}
\Omega_1 & 0 & 0\\
0 & \Omega_2 & 0\\
0 & 0 & \Omega_3
\end{array} } \right) \hspace{1cm} 
M_a = \left({\begin{array}{ccc}
0 & \tau_3 & - \tau_2\\
- \tau_3 & 0 & \tau_1\\
\tau_2 & - \tau_1 & 0
\end{array} } \right)
\end{align}

The equation we get on applying (\ref{fld}) and (\ref{mat}) to (\ref{nahm}) is:

\begin{equation}
\dot{M} = \big( \text{Adj} (M) \big)^T + M^T M - \text{Tr} (M) . M
\end{equation}

and taking the diagonal parts gives us the generalized Darboux-Halphen system equations.

\begin{equation}
\label{gendh} \dot{\Omega}_i = \Omega_j \Omega_k - \Omega_i \big( \Omega_j + \Omega_k \big) + \tau^2 \hspace{2cm} \tau^2 = \tau_1^2 + \tau_2^2 + \tau_3^2
\end{equation}

The off-diagonal terms taken together give us

\begin{equation}
\label{teq} \dot{\tau}_i = - \tau_i \big( \Omega_j + \Omega_k \big) \hspace{2cm} \tau_i^2 = \alpha_i^2 \big( \Omega_j - \Omega_i \big) \big( \Omega_i - \Omega_k \big)
\end{equation}

Now we shall consider equations that will hold for both, classical and generalized systems and their general solutions.

\subsection{Solutions of the generalized system}

While the generalized system has different set of equations from the classical version due to the common extra term $\tau^2$, if we choose to designate the variable as

\vspace{-0.25cm}
\begin{equation}
x_i = \Omega_j - \Omega_k
\end{equation}

we should obtain equations similar to \cite{dhsstruc}. Then, using the generalized Darboux-Halphen equations, we can write

\vspace{-0.5cm}
\begin{align} \displaybreak[0]
\label{drv1} \dot{x}_i = \dot{\Omega}_j - \dot{\Omega}_k = - 2 \Omega_i x_i \hspace{1cm} &\Rightarrow \hspace{1cm}  \Omega_i = - \frac12 \big[ \ln x_i \big]' \\ \nonumber \\
\label{drv2} \Rightarrow \hspace{1cm} \bigg[ \ln \bigg( \frac{x_j}{x_i} \bigg) \bigg]' = \bigg[ \ln &\bigg( - \frac{x_j}{x_i} \bigg) \bigg]' = 2 x_k
\end{align}

This equation applies to both, generalized and classical Darboux-Halphen systems. If we choose to define a variable $s$ as:

\begin{equation}
s = - \frac{x_2}{x_1} \hspace{1cm} \Rightarrow \hspace{1cm} s - 1 = - \frac{x_2 + x_1}{x_1} = \frac{\Omega_1 - \Omega_2}{\Omega_2 - \Omega_3} = \frac{x_3}{x_1}
\end{equation}

then using (\ref{drv1}) and (\ref{drv2}), we should find as in \cite{fintgradflo, fintgdh} that

\[ \frac{\dot{s}}{s \big( s - 1 \big)} = 2 x_1 \hspace{1cm} \frac{\dot{s}}{s - 1} = - 2 x_2 \hspace{1cm} \frac{\dot{s}}{s} = 2 x_3 \]

\begin{equation}
\label{hw1}
\boxed{\Omega_1 = - \frac12 \bigg[ \ln \bigg( \frac{\dot{s}}{s \big( s - 1 \big)} \bigg) \bigg]' \hspace{0.75cm} \Omega_2 = - \frac12 \bigg[ \ln \bigg( \frac{\dot{s}}{s - 1} \bigg) \bigg]' \hspace{0.75cm} \Omega_3 = - \frac12 \bigg[ \ln \bigg( \frac{\dot{s}}{s} \bigg) \bigg]'}
\end{equation}

Now the off-diagonal anti-symmetric terms, recalling (\ref{teq}), give rise to equations:

\[ \frac{\dot{\tau}_1}{\tau_1} = \frac12 \bigg[ \ln \bigg( \frac{\dot{s}^2}{s ( s - 1 )} \bigg) \bigg]' \hspace{0.5cm}
\frac{\dot{\tau}_2}{\tau_2} = \frac12 \bigg[ \ln \bigg( \frac{\dot{s}^2}{s^2 ( s - 1 )} \bigg) \bigg]' \hspace{0.5cm}
\frac{\dot{\tau}_3}{\tau_3} = \frac12 \bigg[ \ln \bigg( \frac{\dot{s}^2}{s ( s - 1 )^2} \bigg) \bigg]' \]


with the following solutions:

\begin{equation}
\label{hw2}
\boxed{\tau_1 = \kappa_1 \frac{\dot{s}}{\sqrt{s ( s - 1 )}} \hspace{1cm} \tau_2 = \kappa_2 \frac{\dot{s}}{s \sqrt{( s - 1 )}} \hspace{1cm} \tau_3 = \kappa_3 \frac{\dot{s}}{\sqrt{s} ( s - 1 )} }
\end{equation}

where $s$ satisfies the Schwarzian equation given by

\vspace{-0.5cm}
\begin{equation}
\begin{split}
\big\{ s, t \big\} + \frac{\dot{s}^2}2 V (s) &= 0 \\ \\
\text{where} \hspace{0.75cm} \big\{ s, t \big\} := \frac{d \ }{dt} \bigg( \frac{\ddot{s}}{\dot{s}} \bigg) - \frac12 \bigg( \frac{\ddot{s}}{\dot{s}} \bigg)^2 \hspace{1cm} V (s) &= \frac{1 - 4 \kappa_2^4}{s^2} + \frac{1 - 4 \kappa_3^4}{(s - 1)^2} + \frac{ \kappa_2^4 + \kappa_3^4 - \kappa_1^4 - 1}{s(s - 1)}
\end{split}
\end{equation}

Thus, if we concentrate only on the diagonal terms, we are able to express the metric related to the classical Darboux-Halphen system.

\subsection{Non-existence of a metric for the generalized system}

Naturally, one can suspect that the classical Darboux-Halphen system and consequently the Bianchi-IX metric is the result of setting $\tau_i = 0$. For the classical system, we should have the metric co-efficients given by the diagonal matrix

\begin{equation}
\label{coef} h_{class} = \left({\begin{array}{cccc}
\Omega_1 \Omega_2 \Omega_3 & 0 & 0 & 0\\
0 & \dfrac{\Omega_2 \Omega_3}{\Omega_1} & 0 & 0\\
0 & 0 & \dfrac{\Omega_3 \Omega_1}{\Omega_2} & 0\\
0 & 0 & 0 & \dfrac{\Omega_1 \Omega_2}{\Omega_3}
\end{array} } \right)
\end{equation}

Now, we notice that the matrix describing the metric $h$ can be given by

\vspace{-0.5cm}
\begin{equation}
\label{matlat} h_{class} = M_{class}^{-1} \ \text{Adj} \big( M_{class} \big) \hspace{2cm} \text{where } \hspace{0.75cm} M_{class} = \left({\begin{array}{cccc}
1 & 0 & 0 & 0\\
0 & \Omega_1 & 0 & 0\\
0 & 0 & \Omega_2 & 0\\
0 & 0 & 0 & \Omega_3
\end{array} } \right)
\end{equation}

where $M_{class}$ is the matrix that produces the classical Darboux-Halphen system. With this in mind, we see that the generalized Darboux-Halphen system (\ref{gendh}) seems to arise from a matrix $M_{gen}$ given as

\begin{equation}
M_{gen} = \left({\begin{array}{cccc}
1 & 0 & 0 & 0\\
0 & \Omega_1 & \tau_3 & - \tau_2\\
0 & - \tau_3 & \Omega_2 & \tau_1\\
0 & \tau_2 & - \tau_1 & \Omega_3
\end{array} } \right)
\end{equation}

However, there are not always a vierbeins or metric counterparts for gauge fields, as there are for curvature and connection components. This shall be elaborated further as follows. \\

The torsion-free form of the 1st Cartan structure equation is

\vspace{-0.25cm}
\begin{equation}
d e^i = - {\omega^i}_j \wedge e^j
\end{equation}

Further examination reveals that

\vspace{-0.25cm}
\[ \partial_\mu {e^i}_\nu \ dx^\mu \wedge dx^\nu = - \omega^i_{\mu j}  {e^j}_\nu \ dx^\mu \wedge dx^\nu \]

\vspace{-0.25cm}
\begin{equation}
\therefore \hspace{1cm} {E_j}^\nu \partial_\mu {e^i}_\nu = - \omega^i_{\mu j}
\end{equation}

Recalling that the spin connections can be expanded as shown below, we can say that

\vspace{-0.25cm}
\begin{align} \displaybreak[0]
\omega_{ij} = \eta^{(+)a}_{ij} A^{(+)a} &+ \eta^{(-)a}_{ij} A^{(-)a} \hspace{2cm} A^{(\pm)a} = A^{(\pm)a}_\mu dx^\mu \\ \nonumber \\
\therefore \hspace{1cm} {E_j}^\nu \partial_\mu {e^i}_\nu &= - \Big[ \eta^{(+)a}_{ij} A^{(+)a}_\mu + \eta^{(-)a}_{ij} A^{(-)a}_\mu \Big]
\end{align}

Now, we can obtain the individual SU(2)$_\pm$ gauge potential function components $A^{(\pm)}_\mu$ in terms of the vierbein components as follows

\vspace{-0.25cm}
\begin{equation}
A^{(\pm)a}_\mu = - \frac14 \eta^{(\pm)a} {E_j}^\nu \partial_\mu {e^i}_\nu
\end{equation}

Thus, if we start with a metric or equivalently the vierbeins, we should be able to get the spin-connections and hence gauge fields, and from there go backwards, however, the opposite is not always possible. 


Since the generalized Darboux-Halphen system is primarily a product of the reduction of the SDYM gauge fields, it may not always be possible to find a metric or its vierbeins that are related to it. The classical Darboux-Halphen system is a special case where $\tau_i = 0 \hspace{0.25cm} \forall \hspace{0.25cm} i$, for which we have the self-dual Bianchi-IX metric (gravitational instanton).

\numberwithin{equation}{section}

\section{Aspects of Flow equations}

Geometric flows describe the evolution of a metric on a Riemannian manifold along the path parameter, under a general non-linear equation, given a symmetric tensor $S_{ij}$ \cite{flowapp, BBLP}. Usually, a system that exhibits geometric flows satisfies the equation

\vspace{-0.25cm}
\begin{equation}
\frac{d g_{ij}}{d \tau} = S_{ij}
\end{equation}

where $S_{ij}$ is symmetric. Some systems exhibit a particular category of such flows known as Ricci flow for which $S_{ij} = - R_{ij}$. Such systems that describe Ricci flows do not preserve volume elements, which are described by the equation:

\vspace{-0.25cm}
\begin{equation}
\frac{d g_{ij}}{d \tau} = - R_{ij}
\end{equation}

The Ricci flow equation introduced by Richard Hamilton in 1982 was a primary tool in Grigory Perelman's proof of Thurston's geometrization conjecture, Poincare conjecture being a special case of that. Ricci flow exhibits many similarities with the heat equation: it gives manifolds more uniform geometry and smooths out irregularities and has proven to be a very useful tool in understanding the topology of arbitrary Riemannian manifolds.

Now, we have already shown that Darboux-Halphen systems are Ricci-flat which means that it is a fixed point of the Ricci flow, usually exhibited by gravitational instantons which are extremal points of the Euclidean Einstein-Hilbert action \cite{HSW}. Looking at the Darboux-Halphen equations, we can see that for the Bianchi-IX metric

\vspace{-0.5cm}
\[ \begin{split}
\frac{d \big( c_i^2 \big) }{d \tau \ \ } = c_i^2 \big( &c_j^2 + c_k^2 - c_i^2 - 2 c_j c_k \big) = c_i^2 \big[ \big( c_j - c_k \big)^2 - c_i^2 \big] \\
\frac{d \big( c_0^2 \big) }{d \tau \ \ } = \frac{d \ }{d \tau} &\big( c_1 c_2 c_3 \big)^2 = c_0^2 \bigg\{ c_i^2 + c_j^2 + c_k^2 - 2 \big( c_i c_j + c_j c_k + c_k c_i \big) \bigg\}
\end{split} \]

Thus, we have the following equations:

\begin{equation}
\boxed{\begin{split}
\frac{d \big( c_0^2 \big) }{d \tau \ \ } &= c_0^2 (\vec{x}) \Big[ c_i^2 + c_j^2 + c_k^2 - 2 \big( c_i c_j + c_j c_k + c_k c_i \big) \Big] \\
\frac{d \big( c_i^2 \big)}{d \tau \ \ } &= c_i^2 (\vec{x}) \big[ \big( c_j - c_k \big)^2 - c_i^2 \big]
\end{split}}
\end{equation}

Now if we set $c_0 = 1$ for the co-ordinate rescaling $dt = c_0 (\vec{x}) \ d r$, then we should get

\begin{equation}
\boxed{\begin{split}
c_i^2 + c_j^2 &+ c_k^2 = 2 \big( c_i c_j + c_j c_k + c_k c_i \big) \\
2 \frac{d c_i }{d \tau} &= \frac1{c_j c_k} \big[ \big( c_j - c_k \big)^2 - c_i^2 \big]
\end{split}}
\end{equation}

which matches and re-confirms the results obtained in \cite{BBLP} where the Ricci tensor for such Bianchi-IX geometry is given by the RHS of the above equation, showing that it does exhibit Ricci flow. For a more general Darboux-Halphen system, the result would be of the form:

\vspace{-0.25cm}
\[ \frac{d \big( c_i^2 \big) }{d \tau \ \ } = \ c_i^2 \Big[ c_j^2 + c_k^2 - c_i^2 - 2 \big( \beta_{ij} \ c_i c_j + \beta_{jk} \ c_j c_k + \beta_{ki} \ c_k c_i \big) \Big] \]

where \ \ $2 \beta_{ij} = \lambda_{jk} - \lambda_{ik}, \hspace{0.5cm} 2 \beta_{jk} = \lambda_{ji} + \lambda_{ki}, \hspace{0.5cm} 2 \beta_{ki} = \lambda_{kj} - \lambda_{ij}$

\[ \text{and} \hspace{1cm} \frac{d \big( c_0^2 \big) }{d \tau \ \ } = \frac{d \ }{d \tau} \big( c_1 c_2 c_3 \big)^2 = c_0^2 (\vec{x}) \big[ c_i^2 + c_j^2 + c_k^2 - 2 \big( \alpha_{ij} \ c_i c_j + \alpha_{jk} \ c_j c_k + \alpha_{ki} \ c_k c_i \big) \big] \]

where $2 \alpha_{ij} = \lambda_{ik} + \lambda_{jk}, \hspace{0.5cm} 2 \alpha_{jk} = \lambda_{ji} + \lambda_{ki}, \hspace{0.5cm} 2 \alpha_{ki} = \lambda_{kj} + \lambda_{ij}$

\begin{equation}
\boxed{\begin{split}
\frac{d \big( c_0^2 \big) }{d \tau \ \ } &= c_0^2 (\vec{x}) \big[ c_i^2 + c_j^2 + c_k^2 - 2 \big( \alpha_{ij} \ c_i c_j + \alpha_{jk} \ c_j c_k + \alpha_{ki} \ c_k c_i \big) \big] \\
\frac{d \big( c_i^2 \big) }{d \tau \ \ } &= c_i^2 (\vec{x}) \big[ c_j^2 + c_k^2 - c_i^2 - 2 \big( \beta_{ij} \ c_i c_j + \beta_{jk} \ c_j c_k + \beta_{ki} \ c_k c_i \big) \big]
\end{split}}
\end{equation}

Thus, Ricci flow is exhibited and implies a self-dual Bianchi-IX metric, otherwise known to be the Darboux-Halphen system describing the evolution of $SU(2)$ 3D geometries.

\section{Other related systems}

The Darboux-Halphen system has analogues and equivalents in various forms of quadratic and non-linear differential equations. In this section, we will describe them in detail.

\subsection{Ramamani to Darboux-Halphen}

The Ramamani system \cite{Ramamani-thesis, Ramamani-paper} is described by the following differential equations

\vspace{-0.5cm}
\begin{equation}
\label{ramamani}
\begin{split}
q\frac{d{\cal P}}{dq} &= \frac{{\cal P}^2 - {\cal Q}}{4} \\
q\frac{d{\tilde {\cal P}}}{dq} &= \frac{{\tilde {\cal P}}{\cal P} - {\cal Q}}{2} \\
q\frac{d{\cal Q}}{dq} &= {\cal P}{\cal Q} - {\tilde {\cal P}}{\cal Q}
\end{split}
\end{equation}

In a recent paper Ablowitz et al. \cite{ach} showed that  Ramamani's system  of differential equations is equivalent to a third order scalar nonlinear ODE found by Bureau \cite{bu},  whose solutions are given implicitly by a Schwarzian triangle function. Under a suitable variable transformation, the Ramamani system produces the classical Darboux-Halphen system. \\

The Ramamani system (\ref{ramamani}) for $q = \dfrac1{2 i \pi}$, is described by the equations

\vspace{-0.5cm}
\begin{equation}
\label{ramamani2} 
\begin{split}
\dot{\mathcal{P}} &= \frac{i \pi}2 \big( {\mathcal{P}}^2 - \mathcal{Q} \big) \\
\dot{\widetilde{\mathcal{P}}} &= i \pi \big( \mathcal{P} \widetilde{\mathcal{P}} - \mathcal{Q} \big) \\
\dot{\mathcal{Q}} &= 2 i \pi \big( \mathcal{P} - \widetilde{\mathcal{P}} \big) \mathcal{Q}
\end{split}
\end{equation}

We convert to Darboux-Halphen variables $\big( {\mathcal P}, \widetilde{{\mathcal P}}, {\mathcal Q} \big) \rightarrow \big( X, Y, Z \big)$ \cite{yurisimonalexey} as follows

\begin{equation}
{\mathcal P} = \frac2{i \pi} X \hspace{1cm} \widetilde{{\mathcal P}} = \frac1{i \pi} \big( 2 X - Y - Z \big) \hspace{1cm} {\mathcal Q} = \frac4{\pi^2} \big( Z - X \big) \big( X - Y \big)
\end{equation}

Naturally, if we apply the above transformation to the Ramamani equations (\ref{ramamani2}), we shall get the classical Darboux-Halphen system equations. 

\vspace{-0.5cm}
\[ \begin{split}
\dot{\mathcal P} = \frac2{i \pi} \dot{X} &= \frac{i \pi}2 \bigg\{ - \frac4{\pi^2} X^2 - \frac4{\pi^2} \big( XY + XZ - YZ - X^2 \big) \bigg\} \\
\Rightarrow \hspace{1cm} &- \frac4{\pi^2} \dot{X} = - \frac4{\pi^2} \big( XY + XZ - YZ \big)
\end{split} \]

and hence, we get one Darboux-Halphen equation in the form

\begin{equation}
\dot{X} = X \big( Y + Z \big) - YZ
\end{equation}

For the others, the process is more elaborate although quite straightforward to show.

\vspace{-0.5cm}
\[ \begin{split}
\dot{\widetilde{\mathcal P}} = \frac1{i \pi} \big( 2 \dot{X} - \dot{Y} - \dot{Z} \big) &= i \pi \bigg\{ - \frac4{\pi^2} X^2 + \frac2{\pi^2} X \big( Y + Z \big) - \frac4{\pi^2} \big( XY + XZ - YZ - X^2 \big) \bigg\} \\
\Rightarrow \hspace{1cm} - \frac1{\pi^2} \big( 2 \dot{X} - \dot{Y} - \dot{Z} \big) &= \frac2{\pi^2} \big( \dot{X} + YZ \big) - \frac4{\pi^2} \dot{X}
\end{split} \]

\begin{equation}
\label{dhtype2} \Rightarrow \hspace{1cm} \dot{Y} + \dot{Z} = 2 YZ
\end{equation}

\[ \begin{split}
\dot{\mathcal Q} = \frac4{\pi^2} \big[ \big( \dot{Z} - \dot{X} \big) \big( X &- Y \big) + \big( Z - X \big) \big( \dot{X} - \dot{Y} \big) \big] = \frac8{\pi^2} \big( Y + Z \big) \big(Z - X \big) \big( X - Y \big) \\
\Rightarrow \hspace{1cm} \big( \dot{Z} - Z^2 &\big) \big( X - Y \big) + \big( Y^2 - \dot{Y} \big) \big( Z - X \big) = 0
\end{split} \]

Using (\ref{dhtype2}), we have another DH equation

\begin{equation}
\dot{Y} = Y \big( Z + X \big) - ZX
\end{equation}

Naturally, using (\ref{dhtype2}) again, we should get the final equation

\begin{equation}
\dot{Z} = Z \big( X + Y \big) - XY
\end{equation}

Consequently, we have three sets of equations for the classical Darboux-Halphen system

\[ \boxed{\begin{split}
\dot{X} = X \big( Y + Z \big) - YZ \\
\dot{Y} = Y \big( Z + X \big) - ZX \\
\dot{Z} = Z \big( X + Y \big) - XY
\end{split}} \]

This is henceforth, another system related to the Ramanujan equations \cite{Ramanujan-arith, Ramanujan-collect} via the focal point of classical DH systems they converge to. We must note that, the Ramamani system gives rise to self dual (and not anti-self dual) Darboux-Halphen equations. Inverting the sign of the Halphen variables gives the familiar anti-self-dual system.

\subsection{The Chazy equation}

Now, we shall see how the solution of the Darboux-Halphen system satisfies the Chazy equation \cite{Chazy, qdschzeq}. Let us take the previous result for anti-self-duality and $\lambda_{ij} = -1$ and write it for all values of $i, j, k$

\vspace{-0.5cm}
\begin{equation}
\begin{split}
\dot{\Omega}_1 = \Omega_2 \Omega_3 - \Omega_1 \big( \Omega_2 + \Omega_3 \big) \\
\dot{\Omega}_2 = \Omega_3 \Omega_1 - \Omega_2 \big( \Omega_3 + \Omega_1 \big) \\
\dot{\Omega}_3 = \Omega_1 \Omega_2 - \Omega_3 \big( \Omega_1 + \Omega_2 \big) \\
\end{split}
\end{equation}

Adding up will give 

\begin{equation}
\dot{\Omega}_1 + \dot{\Omega}_2 + \dot{\Omega}_3 = - \big( \Omega_1 \Omega_3 + \Omega_2 \Omega_1 + \Omega_3 \Omega_2 \big)
\end{equation}

If we define $y = - 2 \big( \Omega_1 + \Omega_2 + \Omega_3 \big)$, we will have

\vspace{-0.25cm}
\begin{align}
\dot{y} &= 2 \big( \Omega_1 \Omega_3 + \Omega_2 \Omega_1 + \Omega_3 \Omega_2 \big) \\
\ddot{y} &= - 12 \ \Omega_1 \Omega_2 \Omega_3
\end{align}

Thus, the third order derivative will be

\vspace{-0.25cm}
\[ \begin{split}
\dddot{y} &= - 12 \big[ \big\{ \Omega_2 \Omega_3 - \Omega_1 \big( \Omega_2 + \Omega_3 \big) \big\} \Omega_2 \Omega_3 + \Omega_1 \big\{ \Omega_3 \Omega_1 - \Omega_2 \big( \Omega_3 + \Omega_1 \big) \big\} \Omega_3 \\
& \hspace{3cm} + \Omega_1 \Omega_2 \big\{ \Omega_1 \Omega_2 - \Omega_3 \big( \Omega_1 + \Omega_2 \big) \big\} \big] \\
&= 48 \ \Omega_1 \Omega_2 \Omega_3 \big( \Omega_1 + \Omega_2 + \Omega_3 \big) - 12 \ \big( \Omega_2 \Omega_3 + \Omega_3 \Omega_1 + \Omega_1 \Omega_2 \big)^2 \\
&= 2 y \ddot{y} - 3 \dot{y}^2
\end{split} \]

Thus, we obtain the Chazy equation \cite{Chazy}

\begin{equation}
\label{chazy} \boxed{\frac{d^3 y}{dt^3} = 2 y \frac{d ^2 y}{d t^2} - 3 \bigg( \frac{d y}{d t} \bigg)^2}
\end{equation}

\numberwithin{equation}{subsection}

\subsection{The Ramanujan equation}

In case of the classical Chazy system, the Ramanujan equations \cite{Ramanujan-arith, Ramanujan-collect} are given by

\vspace{-0.25cm}
\begin{equation}
\label{ram}
\begin{split}
\dot{P} &= \frac{i \pi}6 \big( P^2 - Q \big) \\
\dot{Q} &= \frac{2 i \pi}3 \big( P Q - R \big) \\
\dot{R} &= i \pi \big( P R - Q^2 \big)
\end{split}
\end{equation}

To understand how they are related to the Chazy equation, we shall examine what they imply systematically. From the first equation of (\ref{ram}), we find that

\begin{equation}
\label{a} Q = P^2 - \frac6{i \pi} \dot{P} \hspace{1cm} \Rightarrow \hspace{1cm} Q = Q ( P, \dot{P} )
\end{equation}

Applying (\ref{a}) to the second eq. of (\ref{ram}), we get

\[ \dot{Q} = \dot{Q} ( P, \dot{P}, \ddot{P} ) = 2 \bigg( P \dot{P} - \frac3{i \pi} \ddot{P} \bigg) \]

\begin{equation}
\label{b} \Rightarrow \hspace{0.5cm} R = R (P, \dot{P}, \ddot{P} ) = PQ - \frac3{2 i \pi} \dot{Q} = P^3 - \frac9{i \pi} P \dot{P} - \frac9{\pi^2} \ddot{P}
\end{equation}

Finally, using result (\ref{b}) in the last equation of (\ref{ram}), we will get

\[ \dot{R} =  i \pi \big( P R - Q^2 \big) = 3 P^2 \dot{P} - \frac9{i \pi} \big( \dot{P}^2 + P \ddot{P} \big) - \frac9{\pi^2} \dddot{P} \]

\begin{equation}
\therefore \hspace{1cm} \dddot{P} + i \pi \big( 3 \dot{P}^2 - 2 P \ddot{P} \big) = 0
\end{equation}

However, we are not there yet. The final step requires us to take advantage of the non-linearity of this equation and write $P = \dfrac{y}{i \pi}$ to arrive at

\begin{equation}
\hspace{1cm} \dddot{y} = 2 y \ddot{y} - 3 \dot{y}^2
\end{equation}

This final result is the classical Chazy equation (\ref{chazy}) we are familiar with.

\subsection{Generalized Chazy System}

If we start instead with the generalized Darboux-Halphen equations and set the co-efficients as $\alpha_1 = \alpha_2 = \alpha_3 = \frac2n$, we will get the corresponding generalized Chazy equation \cite{fintgradflo}.

\begin{equation}
\boxed{\frac{d^3 y}{dt^3} - 2 y \frac{d ^2 y}{d t^2} + 3 \frac{d y}{d t}^2 = \frac4{36 - n^2} \bigg( 6 \frac{d y}{d t} - y^2 \bigg)^2}
\end{equation}

The set of transformations that leads the Ramanujan equations to the above generalized Chazy equation turn out to be:

\vspace{-0.5cm}
\begin{equation}
\label{gram} \begin{split}
\dot{P} &= \frac{i \pi}6 \big( P^2 - Q \big) \\
\dot{Q} &= \frac{2 i \pi}3 \big( P Q - R \big) \\
\dot{R} &= i \pi \bigg[ P R - Q^2 \bigg( 1 - \frac{36}{36 - n^2} \bigg) \bigg]
\end{split}
\end{equation}

The first two equations of (\ref{gram}) are the same as for (\ref{ram}), so the same steps will follow as with (\ref{a}) and (\ref{b}), but for the last step, we will have

\begin{equation}
\dddot{P} + i \pi \big( 3 \dot{P}^2 - 2 P \ddot{P} \big) = \frac{4 i \pi }{36 - n^2} \bigg( P^2 - \frac6{i \pi} \dot{P} \bigg)^2
\end{equation}

Applying the same variable redefinition $P = \dfrac{y}{i \pi}$ as before, we obtain

\begin{equation}
\frac{d^3 y}{dt^3} - 2 y \frac{d ^2 y}{d t^2} + 3 \frac{d y}{d t}^2 = \frac4{36 - n^2} \bigg( 6 \frac{d y}{d t} - y^2 \bigg)^2
\end{equation}

which is exactly the generalized Chazy equation described before.

\numberwithin{equation}{section}

\section{Integrability of the Bianchi-IX}

There are various impositions possible on a 4-dimensional Riemannian metric. It could be K\"ahler or Einstein or even have an anti-self-dual (ASD) Weyl tensor. The venn-diagram below depicts the various possibilities resulting to different field equations in 4-dimensions, where the intersection zones correspond to interesting conditions.

\vspace{-0.25cm}
\begin{center}
\includegraphics[scale=0.65]{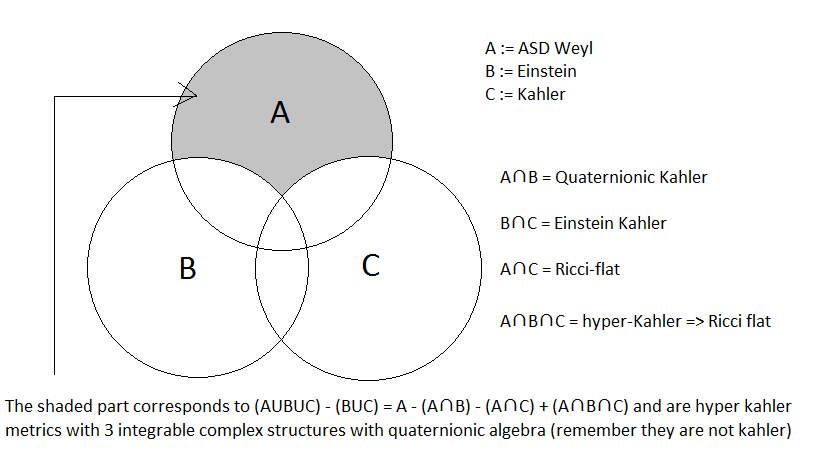}
\end{center}

If ($e^0, e^1, e^2, e^3$) define the vierbeins on a Riemannian 4-manifold, the basis of self-dual 2-forms is given as

\vspace{-0.25cm}
\begin{equation}
* \eta^i = \eta^i = \eta^i_{ab} e^a \wedge e^b:
\begin{cases}
\eta^1 = e^0 \wedge e^1 + e^2 \wedge e^3 \\
\eta^2 = e^0 \wedge e^2 + e^3 \wedge e^1 \\
\eta^3 = e^0 \wedge e^3 + e^1 \wedge e^2 
\end{cases}
\end{equation}

Similarly, the anti-self-dual 2-form basis is given by

\vspace{-0.25cm}
\begin{equation}
* \bar{\eta}^i = - \bar{\eta}^i = \bar{\eta}^i_{ab} e^a \wedge e^b:
\begin{cases}
\bar{\eta}^1 = e^0 \wedge e^1 - e^2 \wedge e^3 \\
\bar{\eta}^2 = e^0 \wedge e^2 - e^3 \wedge e^1 \\
\bar{\eta}^3 = e^0 \wedge e^3 - e^1 \wedge e^2 
\end{cases}
\end{equation}

If ${\omega^i}_j$ are the spin connection 1-forms for the self-dual part, then the first Cartan equation is given by

\vspace{-0.25cm}
\begin{equation}
d \eta^i = {\omega^i}_j \wedge \eta^j
\end{equation}

The curvature 2-form is given as usual by the 2nd Cartan structure equation (\ref{cartan2}). It is possible to expand the curvature in terms of $\eta^i$ and $\bar{\eta}^i$ as

\vspace{-0.25cm}
\begin{equation}
R_{ij} = {W_{ij}}^k \eta_k + {\Phi_{ij}}^k \bar{\eta}_k
\end{equation}

where  $W_{ijk}$ \& $\Phi_{ijk}$ are unknown co-efficients determined by the conditions imposed by various field equations.

\subsection*{Conditions determining $W_{ijk}$ \& $\Phi_{ijk}$}

The 1st Bianchi identity ${R^i}_j \wedge \eta^j = 0$ implies $W_{ijj} = 0$, further implying that $W_{ijk}$ has 6 independent components, out of which 5 correspond to the SD Weyl tensor  and one to the totally anti-symmetric part corresponding to Ricci scalar. On the other hand, $\Phi_{ijk}$ has 9 components corresponding to trace-free Ricci tensor.

\begin{enumerate}

\item Iff $W_{ijk} = \Lambda \epsilon_{ijk}$, where $\Lambda$ is a multiple of Ricci scalar, we are in Set A $\Rightarrow$ ASD Weyl.

\item Iff $W_{ijk} = \Lambda \epsilon_{ijk}$ and $\Phi_{ijk} = 0$, then we have ASD Einstein ($A \cap B$).

\item Iff $W_{ijk} = 0 = \Phi_{ijk}$, we are in $(A \cap B \cap C)$ which is hyper-K\"ahler.

\end{enumerate}

Returning to the Bianchi-IX metric (\ref{bianchiix}) and using the parametrization (\ref{pmtr}), it can be written in terms of the basis $(\sigma^1, \sigma^2, \sigma^3)$  of left-invariant forms on $SU(2)$  as

\begin{equation}
\label{bianchiixnew} ds^2 = \big[ \Omega_1 (r) \Omega_2 (r) \Omega_3 (r) \big] dr^2 + \frac{\Omega_2 \Omega_3}{\Omega_1} \sigma_1^2 + \frac{\Omega_3 \Omega_1}{\Omega_2} \sigma_2^2 + \frac{\Omega_1 \Omega_2}{\Omega_3} \sigma_3^2
\end{equation}

where $\Omega_i, \forall i = 1, 2, 3$ are functions of $r$ and $\sigma^i$s satisfy Maurer Cartan equations. From this form of the metric, the vierbeins can be used to produce the SD 2-forms:

\begin{equation}
\eta^i = \Omega_j \Omega_k \ dr \wedge \sigma^i + \Omega_i \ \sigma^j \wedge \sigma^k \hspace{1cm} i \neq j \neq k
\end{equation}

Therefore, the connection 1-forms $\omega_{ij}$ can be written in terms of arbitrary functions $A_i$ of $r$, $\forall i = 1, 2, 3$ such that 

\vspace{-0.25cm}
\begin{equation}
\omega_{12} = \frac{A_3}{\Omega_3} \sigma^3 + (\text{cyclic permutations})
\end{equation}

All $A_i$ components are obtained from the system below:

\begin{equation}
\label{system1} \dot{\Omega}_i = \Omega_j \Omega_k - \Omega_i \big( A_j + A_k \big) \hspace{2cm} i \neq j \neq k = 1, 2, 3
\end{equation}

We will refer to this system as the first system for future reference. \\

With the help of Cartan calculus, one can find the curvature 2-forms in terms of derivatives of $A_i$s. With the specific choice of field equations from restricitng ourselves to a specific region of the diagram (indicated in the Venn diagram), we will obtain a second system of 1st order differential equations involving $A_i$.

If we  choose regions outside the top circle, we typically get non-integrable equations. Dancer and Strachan \cite{DS} already showed this for Einstein K\"ahler ($B \cap C$), while Barrow \cite{B} showed the same thing for Einstein (Set $B$). However, field equations belonging to the top circle A are integrable, as expected from the heuristic, yet concrete argument that self-duality implies integrability. according to Mason \cite{Ma}. \\

Imposing the vanishing of ASD Weyl tensor and the scalar curvature $\omega_{ij}$ results in the system of the equation widely known as Chazy system \cite{AC, Ch}. This system has a long history, having been studied and solved in the 19th century by Brioschi \cite{Br}.

\begin{equation}
\label{system2} \dot{A}_i = A_j A_k - A_i \big( A_j + A_k \big) \hspace{2cm} i \neq j \neq k = 1, 2, 3
\end{equation}

\newpage
We shall now list the following features of the first and second systems:

\begin{enumerate}

\item[i)] If all $A_i = 0$, then the connection of ASD 2-forms are clearly flat and the metric describes vaccum. This was found by Belinsky \cite{BGPP} and Eguchi-Hanson \cite{EH}.

\item[ii)] If $\Omega_i = A_i, \forall \ i$, then the first and second systems of equations are identical, which is precisely the Atiyah and Hitchin's \cite{AH} case.

\item[iii)] If one insists all $A_i$s to be constant in r, then without any loss of generality, if two of them necessarily vanish, then the remaining $A_i \neq 0$ can reduce the first system to a special case of Painleve-III \cite{Tod}. Also, the form $\eta^i$ is covariant constant in this case so that one arrives at the Pederson-Poon scalar flat K\"ahler metric \cite{PP}.

\item[iv)] There exists a significant conserved quantity

\vspace{-0.25cm}
\begin{equation}
\label{conserved}
Q = \frac{\Omega_1^2}{(A_1 - A_2)(A_1 - A_3)} + \frac{\Omega_2^2}{(A_2 - A_1)(A_2 - A_3)} + \frac{\Omega_3^2}{(A_3 - A_1)(A_3 - A_2)}
\end{equation}

\end{enumerate}

There is a covariance under fractional linear transformations in $r$ \cite{Tod}, which means that the solutions of the second system with $A_1 = A_2 \neq 0$ is conformally related to the Pederson-Poon case \cite{PP}. \\

Now, for a general solution of the second system, we introduce a new dependent variable $x$ as per Brioschi \cite{Br}

\vspace{-0.25cm}
\begin{equation}
x = \frac{A_1 - A_2}{A_3 - A_2}
\end{equation}

It is now straightforward to show that (\ref{system2}) reduces to the 3rd order ODE for $x$

\vspace{-0.25cm}
\begin{equation}
\dddot{x} = \frac32 \frac{\ddot{x}^2}{\dot{x}} - \frac12 \big( \dot{x} \big)^3 \bigg( \frac1{x^2} + \frac1{x (x - 1)} + \frac1{(x - 1)^2} \bigg)
\end{equation}

A remarkable fact is that this ODE is satisfied by the reciprocal of the elliptic modular function. Now this elliptic modular function has a natural boundary in the r-plane, so the $A_i$ and hence $\Omega_i$ have a natural boundary in the r-plane and the location of the boundary depends on the constants of integration. This implies the self-duality equations are not always equivalent to Painleve property, and thus integrable. \\

Now we introduce new dependent variables $\rho_i, \forall  i = 1, 2, 3$ according to

\vspace{-0.25cm}
\begin{equation}
\Omega_1 = \rho_1 \frac{\dot{x}}{\sqrt{ x (1 - x)}} \hspace{1cm}
\Omega_2 = \rho_2 \frac{\dot{x}}{ x \sqrt{(1 - x)}} \hspace{1cm}
\Omega_3 = \rho_3 \frac{\dot{x}}{\sqrt{x} (1 - x)}
\end{equation}

and switch independent variable from $r$ to $x$ (ie. $\dot{x} \equiv \frac{dx}{dr}$), so the first system becomes

\vspace{-0.25cm}
\begin{equation}
\frac{d \rho_1}{dx} = \frac{\rho_2 \rho_3}{x (1 - x)} \hspace{2cm}
\frac{d \rho_2}{dx} = \frac{\rho_3 \rho_1}x \hspace{2cm}
\frac{d \rho_3}{dx} = \frac{\rho_1 \rho_2}{(1 - x)}
\end{equation}

This system is known to reduce to Painleve VI with the first integral

\begin{equation}
\gamma = \frac12 \big( \rho_2^2 + \rho_3^2 - \rho_1^2 \big) = const
\end{equation}

which is in fact the conserved quantity (\ref{conserved}), with the new metric being

\begin{equation}
ds ^2 = \frac{\rho_1 \rho_2 \rho_3}{x (1 - x)} \dot{x} \bigg[ \frac{d x^2}{x (1 - x)} + \frac{(\sigma^1 )^2}{\rho_1^2} + \frac{(1 - x)(\sigma^2 )^2}{\rho_2^2} + \frac{x(\sigma^3 )^2}{\rho_3^2} \bigg]
\end{equation}

Now we shall solve the new version of the first system where we will try to form an equation for $\rho_3$ only. Due to the existence of the first integral $\gamma$, this will be second order equation. To recognize it better, we introduce a new independent variable $z$, given as

\begin{equation}
x = \frac{4 \sqrt{z}}{\big(1 + \sqrt{z} \big)^2}
\end{equation}

a new dependent variable $V$ given by

\begin{equation}
\rho_3 = \frac zV \frac{d V}{d z} - \frac V{2 (z - 1)} - \frac12 + \frac z{2 V (z - 1)}
\end{equation}

It is not very tedious to show that $V$ satisfies Painleve-VI equations with parameters ($\alpha, \beta, \gamma, \delta$) in the notation of \cite{Ince} or ($\frac18, - \frac18, \gamma, \frac12 (1 - 2 \gamma)$) in the notation of \cite{AC} . Thus, we see the equation for the conformal factor

\begin{equation} 
\Theta = \dfrac{\rho_1 \rho_2 \rho_3}{x (1 - x)} \dfrac{d x}{d t}
\end{equation}

 has the Painleve property, but it also contains the function $x (\tau)$ which has a natural boundary. The choice of conformal factor is equivalent to a gauge choice to make the Ricci scalar vanish. \\

Now we have found the general solution of the metric (\ref{bianchiixnew}) inside the top circle of the figure (the shaded region of A). Also note that ASD Bianchi-IX metrics are not always diagonal in the chosen invariant basis of 1-forms. We can always adjust the conformal factor $\Theta$ in order to make this ASD Bianchi type metric to become Einstein. This would constitute a metric for the region $A \cap B$, which are quaternionic K\"ahler type of metrics and are diagonal in the basis.

The solution for the conformal factor was found in \cite{Tod95} with $\gamma = \dfrac18$ and writing down the desired condition as a set of equations on $\Theta$ and finally solve it. After a little algebra and once all the dusts get settled, we get 
$\Theta = \dfrac{N}{D^2}$, with

\vspace{-0.5cm}
\begin{equation}
\begin{split}
N &= 2 \rho_1 \rho_2 \rho_3 ( 4x \rho_1 \rho_2 \rho_3 + P ) \\
P &= x \big( \rho_1^2 + \rho_2^2 \big) - (1 - 4 \rho_3^2 ) \big( \rho_2^2 - (1 - x) \rho_1^2 \big) \\
D &= x \rho_1 \rho_2 + 2 \rho_3 \big( \rho_2^2 - (1 - x) \rho_1^2 \big)
\end{split}
\end{equation}

Since the equation for $\rho_3$ is a 2nd order differential equation, the metric depends on two arbitrary constants. Particularly it is worth mentioning that there exists ASD Einstein metrics on $S^4$, which with appropriate choices of field equations fill in the general left-invariant metric on $S^3$ similar to the case of a 4-dim hyperbolic metric that fills the round metric on $S^3$.

\section{Conclusions}

Summarizing we have seen that there exist two different approaches that lead up to the classical Darboux-Halphen system. One approach starts from the Bianchi-IX metric and considers its anti-self-dual case, while the other starts with a reduction of the self-dual Yang-Mills equation and takes only the diagonal elements of the resulting matrix equation. When we start with SDYM gauge fields, it is clear why we cannot always reliably find a metric or its vierbeins that correspond to the generalized DH system. The classical configuration is a typical prototype where it is possible, as was seen from the way we could obtain it from the self-dual Bianchi-IX metric.

We have computed the form of curvature and confirmed that self-dual cases turn out to be Ricci-flat. We also discovered that the classical Darboux-Halphen is found to exhibit Ricci flow for a modified Bianchi-IX system. We also confirmed that this system satisfies the Chazy equation as well and is strongly related to other systems of differential equations, such as the Ramanujan and Ramamani systems.

It remains a challenge to find other integrable systems of number theoretic importance. Another useful direction could be to solve DH type systems (\ref{hw1}) and (\ref{hw2}) using moving monodromy methods and compare the results. We are also trying to find out interesting 1+1 dimensional or 2+1 dimensional DH type systems that can be solved using inverse scattering transform and can be studied to uncover several widely known aspects of integrability. Another testing ground could be various scalar flat K\"ahler metrics namely the LeBrun metric with
$U(1)$ isometry that contains Gibbons-Hawking, Real heaven and Burns metric as special limits and were used to test the bottom-up approach of emergent gravity \cite{Lee:2012px}. Using monodromy evolving deformation \cite{PRL} on Plebanski type  self dual Einstein equations which are actually the EOM obtained from the 2- dim chiral $U(N)$ model in the large $N$ limit and studying integrability might clarify some important issues for the test of emergent gravity \cite{Lee:2012rb}.

\section{APPENDIX}

\subsection*{t'Hooft symbols}

Matrices representing the t'Hooft symbols would be given by :

\vspace{-0.25cm}
\begin{align}
\eta^{(+)1} = \left({\begin{array}{cccc}
0 & 0 & 0 & 1\\
0 & 0 & 1 & 0\\
0 & -1 & 0 & 0\\
-1 & 0 & 0 & 0
\end{array} } \right) \hspace{0.5cm}
\eta^{(+)2} = \left({\begin{array}{cccc}
0 & 0 & -1 & 0\\
0 & 0 & 0 & 1\\
1 & 0 & 0 & 0\\
0 & -1 & 0 & 0
\end{array} } \right) \hspace{0.5cm}
\eta^{(+)3} = \left({\begin{array}{cccc}
0 & 1 & 0 & 0\\
-1 & 0 & 0 & 0\\
0 & 0 & 0 & 1\\
0 & 0 & -1 & 0
\end{array} } \right)\\
\eta^{(-)1} = \left({\begin{array}{cccc}
0 & 0 & 0 & -1\\
0 & 0 & 1 & 0\\
0 & -1 & 0 & 0\\
1 & 0 & 0 & 0
\end{array} } \right) \hspace{0.5cm}
\eta^{(-)2} = \left({\begin{array}{cccc}
0 & 0 & -1 & 0\\
0 & 0 & 0 & -1\\
1 & 0 & 0 & 0\\
0 & 1 & 0 & 0
\end{array} } \right) \hspace{0.5cm}
\eta^{(-)3} = \left({\begin{array}{cccc}
0 & 1 & 0 & 0\\
-1 & 0 & 0 & 0\\
0 & 0 & 0 & -1\\
0 & 0 & 1 & 0
\end{array} } \right)
\end{align}

which obey the following relations between themselves

\vspace{-0.25cm}
\begin{equation}
\sum_{i = 1}^3 \eta^{(\pm)i}_{\mu \nu} \eta^{(\pm)i}_{\lambda \gamma} = \delta_{\mu \lambda} \delta_{\nu \gamma} - \delta_{\mu \gamma} \delta_{\nu \lambda} \pm \varepsilon_{\mu \nu \lambda \gamma}
\end{equation}

\vspace{-0.25cm}
\begin{align}\big[ \eta^{(\pm) i}, \eta^{(\pm) j} \big]_{\mu \nu} = -2 \epsilon^{ijk} &\eta^{(\pm) k}_{\mu \nu} \\ \nonumber\\
\big[ \eta^{(\pm) i}, \eta^{(\mp) j} \big]_{\mu \nu} = 0 \hspace{0.5cm} \Rightarrow \hspace{0.5cm} &\eta^{(\pm) i}_{\mu \rho} \eta^{(\mp) j}_{\rho \nu} = \eta^{(\pm) j}_{\nu \rho} \eta^{(\mp) i}_{\rho \mu}
\end{align}

\vspace{-0.25cm}
\begin{align}
\big\{ \eta^{(\pm) i}, \eta^{(\pm) j} \big\}_{\mu \nu} = -2 \delta^{ij} &\delta_{\mu \nu} \hspace{2cm} \\ \nonumber\\
\big\{ \eta^{(\pm) i}, \eta^{(\mp) j} \big\}_{\mu \nu} = 0 \hspace{0.5cm} \Rightarrow \hspace{0.5cm} &\eta^{(\pm) i}_{\mu \nu} \eta^{(\mp) j}_{\mu \nu} = 0
\end{align}

\begin{equation}
\therefore \hspace{0.5cm} \eta^{(\pm) i}_{\mu \lambda} \eta^{(\pm) j}_{\nu \lambda} = \delta^{ij} \delta_{\mu \nu} + \epsilon^{ijk} \eta^{(\pm) k}_{\mu \nu}
\end{equation}

\begin{equation}
\epsilon_{\mu \nu \lambda \gamma} \eta^{(\pm)i}_{\gamma \sigma} = \mp \big( \delta_{\sigma \lambda} \eta^{(\pm)i}_{\mu \nu} + \delta_{\sigma \mu} \eta^{(\pm)i}_{\nu \lambda} - \delta_{\sigma \nu} \eta^{(\pm)i}_{\mu \lambda} \big)
\end{equation}

\begin{equation}
\therefore \hspace{0.5cm} \epsilon^{ijk} \eta^{(\pm) j}_{\mu \nu} \eta^{(\pm) k}_{\rho \sigma} = \delta_{\mu \sigma} \eta^{(\pm) i}_{\rho \nu} - \delta_{\nu \sigma} \eta^{(\pm) i}_{\rho \mu} + \delta_{\rho \mu} \eta^{(\pm)i}_{\nu \sigma} - \delta_{\rho \nu} \eta^{(\pm)i}_{\mu \sigma}
\end{equation}

\section*{Acknowledgement}
One of the authors SC would like to thank Marios Petropoulos for several correspondences during the course of this project and Thanu Padmanabhan for an enlightening discussions on a subject related to this article. The research of RR was supported by FAPESP through Instituto de Fisica, Universidade de Sao Paulo with grant number 2013/17765-0.


\begin{thebibliography}{999}


\bibitem{GH78} G. W. Gibbons and S. W. Hawking, {\it  Gravitational Multi-instantons},  Phys. Lett. B 78 (1978).

\bibitem{GH79} G. W. Gibbons and S. W. Hawking, {\it Classification of Gravitational Instanton Symmetries}, Comm. Math. Phys. 66, 291-310 (1979).

\bibitem{GH-book} G. W. Gibbons and S. W. Hawking, {\it Euclidean Quantum Gravity}, World Scientific (1993).

\bibitem{ACH} M.J. Ablowitz, S. Chakravarty and R.G. Halburd, {\it Integrable systems and reductions of the self-dual Yang-Mills equations} (2003)  J. Math. Phys. \textbf{44} 3147-72.

\bibitem{lecture} George F. R. Ellis and Henk van Elst {\it Cosmological models} (Cargese lectures 1998)

\bibitem{MM1} Yuri Manin and Matilde Marcolli {\it Symbolic Dynamics, Modular Curves, and Bianchi IX Cosmologies}, arXiv:1504.04005 [gr-qc].

\bibitem{MM2} Yuri Manin and Matilde Marcolli {\it Big Bang, Blowup, and Modular Curves: Algebraic Geometry in Cosmology}, arXiv:1402.2158 [gr-qc], SIGMA 10 (2014), 073.

\bibitem{MPPV} P. Marios Petropoulos and Pierre Vanhove, {\it Gravity, strings, modular and quasimodular forms}, arXiv:1206.0571 [math-ph], CPHT-RR005.0211, IPHT-t12/016, IHES/P/12/08.

\bibitem{HBC} D. Hobill, A. Burd, A.A. Coley, {\it Deterministic Chaos in General Relativity}, Nato Science Series B:, Vol. 332, Plenum Press New York (1994).

\bibitem{PMPetro1} P.M. Petropoulos, {\it Gravitational instantons, Ricci flows and integrable structures}, Talk during Workshop on
Field Theory and Geometric Flows at LMU, Munich.

\bibitem{PMPetro2} P.M. Petropoulos, {\it Self-duality, gravitational instantons and geometric flows}, Talk during Geometric Flows in Mathematics and Theoretical Physics conference at SNS, Pisa.

\bibitem{GibbonsCQG} M. Cvetic, G. W. Gibbons, H. Lu and C. N. Pope  {\it Bianchi IX self-dual Einstein metrics and singular G2 manifolds}, arXiv:0206151 [hep-th], Class. Quant. Grav. Vol 20, No.19

\bibitem{AC} M.J. Ablowitz and P.A. Clarkson, {\em Solitons, nonlinear equations and inverse scattering}. 1991, LMS lecture note series 149.

\bibitem{GibWar} G. W. Gibbons and C. M. Warnick  {\it Hidden symmetry of hyperbolic monopole motion}, arXiv:0609051 [hep-th], J.Geom.Phys.57 : 2286, 2007

\bibitem{FFM} Wentao Fan, Farzad Fathizadeh and Matilde Marcolli {\it Modular forms in the spectral action of Bianchi IX gravitational instantons}, arXiv:1511.05321 [math.DG].

\bibitem{CGR1} Sumanto Chanda, Partha Guha, Raju Roychowdhury, {\it Emergent Schwarzschild : symplectic, geometric and topological perspective} arXiv:1406.6459v3

\bibitem{CGR2} Sumanto Chanda, Partha Guha, Raju Roychowdhury, {\it Taub-NUT and Dynamical Systems : the geometric connection demystified} arXiv:1503.08183v2.

\bibitem{ETH} Tohru Eguchi and Andrew J. Hanson, {\it Self-Dual Solutions to Euclidean Gravity } Annals of Physics 120.1 (1979): 82-106.


\bibitem{dhsstruc} M.J. Ablowitz, S. Chakravarty and R.G. Halburd, {\it The Darboux-Halphen System and the Singularity Structure of its Solutions}, Proc. 4th Int. Conf. on Mathematical and Numerical Aspects of Wave Propagation. 1998.

\bibitem{Nahm} W. Nahm, {\it All self-dual multimonopoles for arbitrary gauge groups}, CERN, preprint TH. 3172., NATO ASI, B 82 (1983) pp.301

\bibitem{Hitch} N. Hitchin, {\it On the construction of monopoles}. Comm. Math. Phys. 89 (2): 145

\bibitem{Donald} S. Donaldon,{\it Nahm's equations and the classification of monopoles}, Comm. Math. Phy. 96 (3), 387

\bibitem{fintgradflo} S. Chakravarty and R.G. Halburd, {\it First integrals and gradient flow for a generalized Darboux-Halphen system}, Contemp. Math., vol. 301, Amer. Math. Soc., Providence, RI, 2002, 273-281.

\bibitem{fintgdh} S. Chakravarty and R.G. Halburd, {\it First integrals of a generalized Darboux-Halphen system}, J. Math. Phys., Vol. 44, No. 4, April 2003.

\bibitem{flowapp} P.M. Petropoulos, {\it Geometric flows and applications}, CPHT-PC012.0210, arXiv:1011.1106v1 [hep-th].

\bibitem{BBLP} Ioannis Bakas, Francois Bourliot, Dieter Lust, Marios Petropoulos, {\it Geometric flows in Horava-Lifshitz gravity}, JHEP 04 (2010)131, arXiv:1002.0062 [hep-th].

\bibitem{HSW} G. Holzegel, T. Schmelzer, C. Warnick, {\it Ricci Flow of Biaxial Bianchi IX Metrics}, arXiv:0706.1694v1 [hep-th].

\bibitem{Ramamani-thesis}
Ramamani V 1970 On some identities conjectured by Ramanujan in his
lithographed notes connected with partition theory and elliptic modular
functions--their proofs--interconnection with various other topics in the
theory of numbers and some generalizations thereon {\em Ph.D. thesis},
The university of Mysore 

\bibitem{Ramamani-paper}
Ramamani V 1989 On some algebraic identities connected with Ramanujan's work
{\em Ramanujan International Symposium on Analysis} ed N K Thakare 
(New Delhi: Macmillan India) pp 279--91

\bibitem{ach}  M. Ablowitz, S. Chakravarty and H. Hahn, {\em Integrable systems and modular forms of level $2$},  J. Phys. A: Math. Gen. 39  (2006),  no. 50, 15341--15353.

\bibitem{bu} F. Bureau, {\em Differential equations with fixed critical points}, Annali di Matematica 64 (1964) 229 - 364.

\bibitem{yurisimonalexey} Yurii V. Brezhnev, Simon L. Lyakhovich and Alexey A.Sharapov, {\em Dynamical systems determining Jacobi's $\theta$-constants}.  Journ. Math. Phys. (2011), 52(11), 112704 (1-21)

\bibitem{Ramanujan-arith}
Ramanujan S 1916 On certain arithmetical functions {\em Trans. Cambridge Philos. Soc.}
\textbf{22} 159--84

\bibitem{Ramanujan-collect}
Ramanujan S 1927 \emph{Collected Papers} (Cambridge: Cambridge University Press);
reprinted 1962 (New York: Chelsea); reprinted 2000 (Providence: American Mathematical Society)

\bibitem{Chazy}
Chazy J 1909 Sur les \'equations diff\'erentielles dont l'int\'egrale
g\'en\'erale est uniforme et admet des singularit\'es essentielles mobiles
{\em C. R. Acad. Sci. Paris} \textbf{149} 563--5

\bibitem{qdschzeq} Robert S. Maier, {\it Quadratic differential systems and Chazy equations 1}, arXiv:1203.0283v2 [math.CA].

\bibitem{DS} Andrew S. Dancer and Ian A.B. Strachan, {\em Einstein metrics on tangent bundles of spheres}.  1994. Math. Proc. Camb. Phil. Soc. 115, (513-525).

\bibitem{B} John D. Barrow, {\em General relativistic chaos and nonlinear dynamics}. 1982, Gen. Rel. Grav. 14 523-530.

\bibitem{Ma} L.J. Mason, 1990, {\em Further advances in Twistor Theory}.

\bibitem{Ch} J. Chazy {\em General relativistic chaos and nonlinear dynamics}. 1910, C.R. Acad. Sci 150, 456.

\bibitem{Br} F. Brioschi, 1881, C.R. Acad. Sci. tXCII 1389.

\bibitem{BGPP} V.A. Belinsky, G.W. Gibbons, D.N. Page, C.N. Pope, {\em Asymptotically Euclidean Bianchi IX metrics in quantum gravity}, Phys. Lett. B, 76 (1978) 433.

\bibitem{EH} Tohru Eguchi and Andrew J. Hanson, {\em Asymptotically flat self-dual solutions to euclidean gravity}, Phys. Lett. B, 237 (1978) 249.

\bibitem{AH} M.F. Atiyah and N.J. Hitchin, {\em The geometry and dynamics of magnetic monopoles}, (Princeton University Press, 1988), Phys. Lett. A, 167 (1985) 21.

\bibitem{Tod} K.P. Tod, {\em A comment on a paper of Pederson and Poon (general relativity)}, 1991 Class. Quant. Grav. 8, 1049-1051.

\bibitem{PP} H. Pedersen and Y.S. Poon, {\em K\"ahler surfaces with zero scalar curvature}, 1990 Class. Quant. Grav. 7, 1707.

\bibitem{Ince} Ince, E.L., {\em Ordinary Differential Equations}, 1956, Dover.

\bibitem{Tod95} K.P. Tod, 1995, {\em Twistor Theory}, Ed. S. Huggett, Dakker, Lecture Notes in Pure and Applied Mathematics 169.


\bibitem{Lee:2012px} S.Lee, R.Roychowdhury and H.S.Yang, {\it Notes on Emergent Gravity}, JHEP 1209, 030 (2012), arXiv:1206.0678 [hep-th].

\bibitem{PRL} M.J. Ablowitz and S. Chakravarty, {\it Integrability, Monodromy Evolving Deformations, and Self-Dual Bianchi IX Systems}, Phys. Rev. Lett.76. No.6, 857 (1996).

\bibitem{Lee:2012rb} S.Lee, R.Roychowdhury and H.S.Yang, {\it Test of Emergent Gravity}, Phys.Rev.D 88, 086007 (2013), arXiv:1211.0207 [hep-th].


\end{thebibliography}
\end{document}